# Multi-spectroscopic investigation of the structure of single-wall carbon nanotubes

Nicolas Izard[1,2], Didier Riehl[1] and Eric Anglaret[2]

1-DGA/DCE/CTA/LOT, 16, bis Avenue Prieur de la Côte d'Or, 94114 Arcueil Cedex, France
2-GDPC, Université Montpellier II, Place E.Bataillon, 34095 Montpellier Cedex 5, France

**Abstract.** We present a multispectroscopic structural study of various nanotube samples with different tube diameters. We determine for each sample the mean bundle and tube diameter as well as the tube diameter distribution. The possibility to work on SWNT of various structural characteristics opens new opportunities to correlate the nanotube structure and their physical properties.

## INTRODUCTION

Single-Wall Carbon Nanotubes (SWNTs) display unique electronic properties due to their monodimensionality. Their electronic properties as well as their linear and non-linear optical properties directly depend on their structure. In particular, for a better understanding of the optical limiting properties of SWNTs, detailed studies of their structure and electronic properties are required [1].

In these proceedings, we present a multispectroscopic study of SWNT samples produced by different techniques. All samples were studied by scanning and high-resolution transmission electron microscopy, X-ray diffraction, Raman spectroscopy and linear optical absorption. The correlations obtained between these different techniques will be emphasized.

## EXPERIMENTAL

SWNTs samples were provided by commercial sources. Three kinds of nanotubes were studied: i) SWNTs synthesized by the electric arc process [2] and purified by a multi-step acid treatment, ii) SWNTs produced by pulverisation laser vaporisation, using a double-pulse laser [3], iii) SWNTs synthesized by catalytic decomposition with the HiPCO process [4]. In this paper, these samples will be refered as Electric Arc (EA), Laser and HiPCO, respectively.

Scanning electron microscopy (SEM) images were recorded with a field emission microscope JEOL JSM 6300F. High-resolution transmission electron microscopy (HRTEM) experiments were performed on a 200 kV Philips CM20 microscope. X-ray diffraction data (XRD) were collected using the $K_\alpha$ radiation of a Cu source ($\lambda$=1.542 Å) and a curve position sensitive detector (INEL-CPS 120). Raman spectra were measured on a dispersive Jobin-Yvon T64000 spectrometer using the 488, 514.5 and

647.1 nm lines of an Ar/Kr laser, and on a FT Bruker RFS100 using the fundamental laser line of a Nd:YAG laser at 1064 nm. Optical absorption spectra were recorded with an UV-Visible-NIR Cary 500 spectrophotometer. For these latter measurements, SWNTs were first dispersed in ethanol and then pulverized and dried on an optical glass substrate.

## RESULTS

SEM and HRTEM pictures typical of each sample are presented in Fig. 1. From SEM pictures, one remarks that EA and HiPCO SWNTs are of very good purity. Laser sample is also of good purity, but small amounts of carbon nanoparticles can be observed. On HRTEM pictures, one remarks that for all samples, tubes assemble into crystalline bundle structures. Note also that at this magnification, nanometric-size catalyst particles are observed on the surface of HiPCO bundles. The mean bundle diameter was estimated from the SEM pictures and the results are reported in Table 1. SWNTs diameter was also estimated from several HRTEM pictures. An FFT-based spatial image filtering was used to measure the periodicity of the SWNTs bundle lattice. The mean-diameter of the nanotubes was estimated from the lattice periodicity, assuming that SWNTs form a hexagonal close-packed structure.

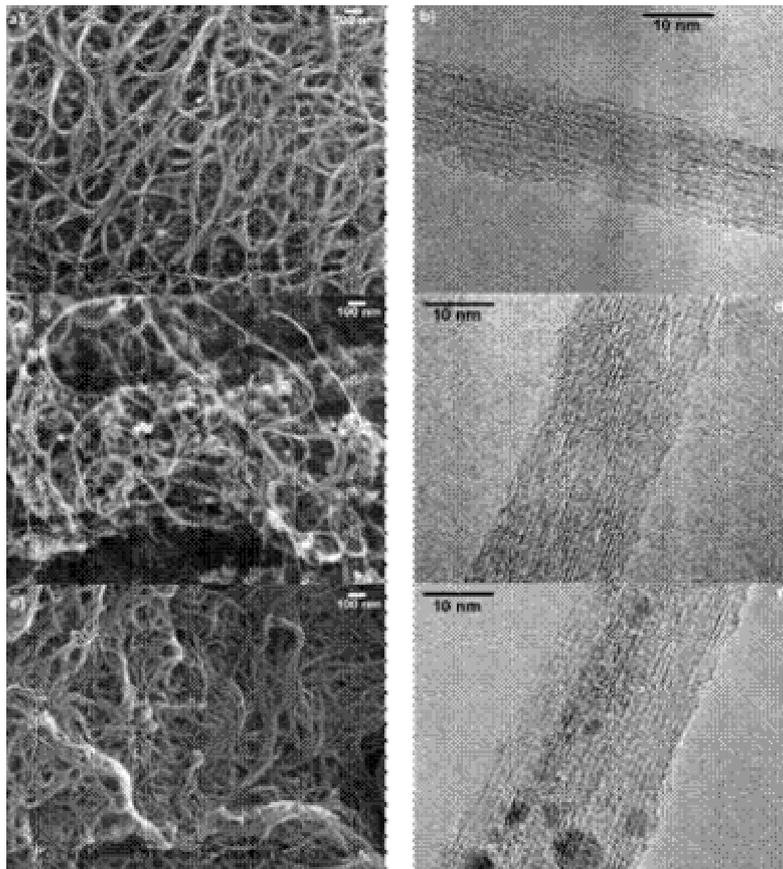

**FIGURE 1.** Left : SEM (50 k magnification), Right : HRTEM (250 k magnification). From top to bottom : EA, Laser, HiPCO samples.

**TABLE 1.** Bundle and tube diameters as estimated from different techniques

| Bundle diameter | EA | Laser | HiPCO |
|---|---|---|---|
| SEM | 20 – 30 nm | 15 – 22 nm | 7 – 12 nm |

| Bundle diameter | EA | Laser | HiPCO |
|---|---|---|---|
| Diameter (HRTEM) | 1.23 – 1.42 nm | 1.34 ± 0.15 nm | 0.9 – 1.3 nm |
| Diameter (XRD) | 1.38 ± 0.15 nm | - | 0.6 – 1.2 nm |
| Diameter (Raman) | 1.39 ± 0.16 nm | 1,34 ± 0.15 nm | 0.85 – 1.36 nm |
| Diameter Range (OA) | 1.1 – 1.5 nm | - | 0.7 – 1.4 nm |

Typical X-Ray diffraction data are presented in Fig. 2. Several points of interest can be noted. First, there is a sharp increase of intensity at very low Q, due to the form factor of individual nanotubes. Second, a series of peaks is observed between 0.4 Å$^{-1}$ and 1.8 Å$^{-1}$, which corresponds to diffraction on the bundle lattice. The most intense peak is the (10) peak, it is measured around 0.42 Å$^{-1}$ for EA and laser samples and around 0.5 Å$^{-1}$ for HiPCO samples. Lastly, the signatures of carbon impurities (graphite and carbon nanoparticles) are observed around 1.8 Å$^{-1}$ for EA and laser samples. Experimental data are fitted using a simple X-Ray diffraction model, which assumes that SWNTs are continuously charged infinite-cylinders, closely packed in a finite-sized bundle, and that each bundle is formed by a set of similar tubes [5]. To calculate the diffraction pattern, we summed the intensity scattered by a set of bundles assuming gaussian distributions of the tube diameter. The distribution that provides the best fit is presented in the insets. For the EA sample, a good fit can be achieved for the entire pattern (Fig. 2, left). The data and fit are very close for laser samples (not shown). For HiPCO sample, the fit is more problematic because of the weakness of the signal and of the broadness of the peaks. The distribution of tube diameter is known to be rather large for these samples [4,7]. No good fits could be achieved using a single (monomodal) gaussian distribution, except if the maximum of the gaussian is shifted down to 6 Å (Fig 2, middle). Therefore, we attempted to fit the data using the sum of two gaussian distributions (bimodal distribution). The best fit is presented in the right part of Fig. 2. However, the fit is not selective enough to conclude definitely. Note that a rather poor fit was achieved with the bimodal distribution proposed in ref. [7]. In addition, the broad band centered at 1.5 Å$^{-1}$ is not well fitted in our model. This is due to a not-corrected fluorescence contribution, and maybe also to the limits of our model, which considers monodisperse tubes in each bundle.

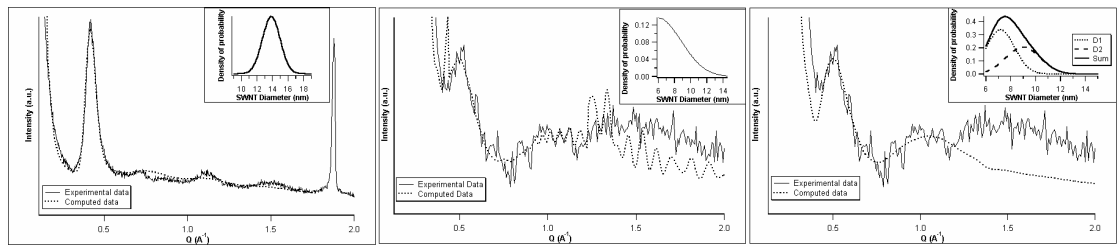

**FIGURE 2.** XRD data (solid line) and fit (dotted line). Left: EA sample; Middle: HiPCO sample, single gaussian distribution. Right: same HiPCO sample, sum of two gaussian distributions. The diameter distributions are plotted in the insets.

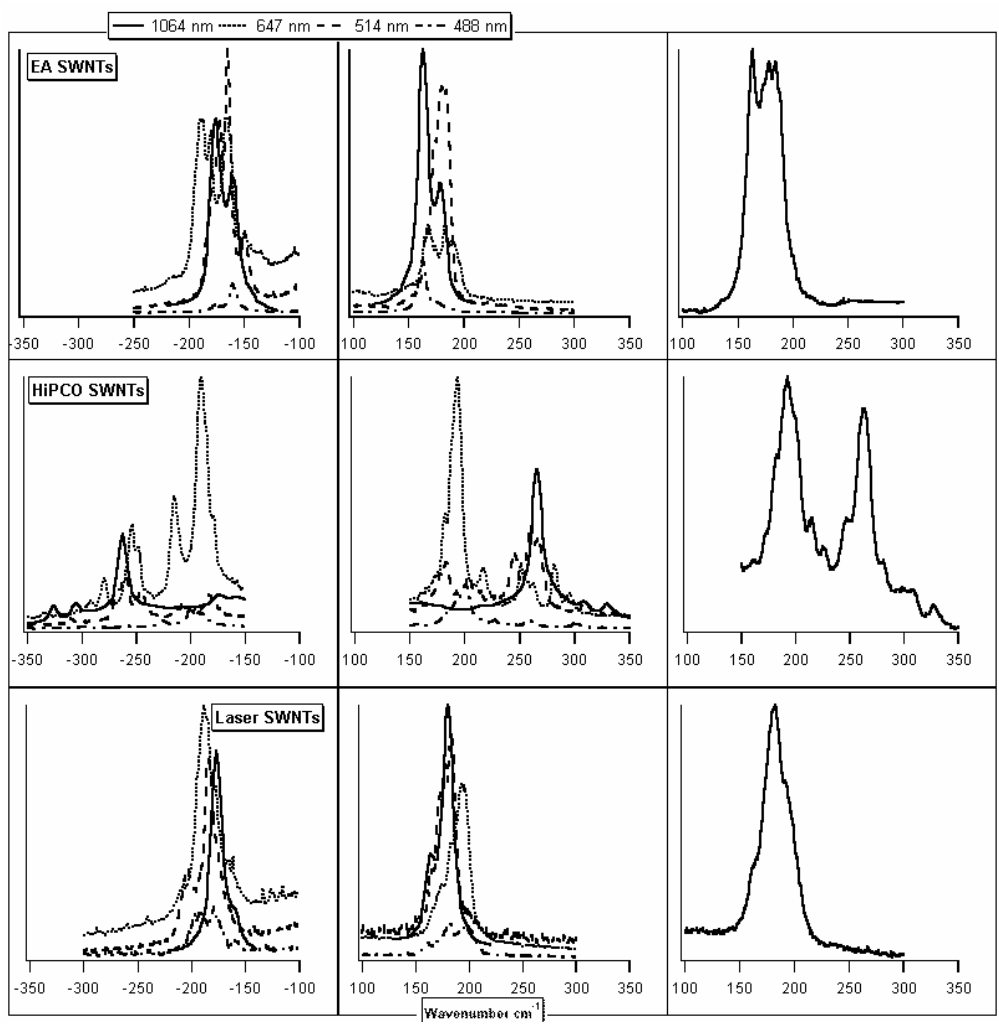

**FIGURE 3.** Left: Anti-Stokes Raman spectra. Middle: Stokes Raman spectra. Right: Normalized sum of Stokes plus Anti-Stokes Raman spectra.

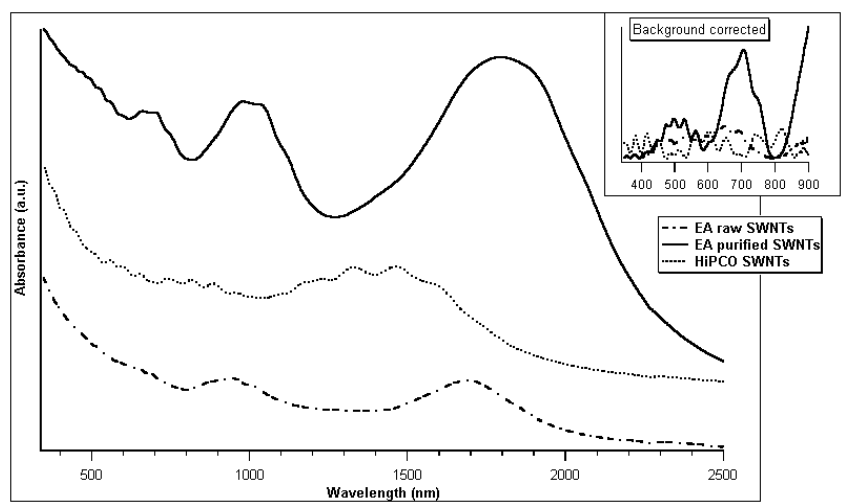

**FIGURE 4.** Optical absorption spectra of SWNTs. Inset: Background corrected.

Fig. 3 displays Raman spectra for all samples in the frequency range of the radial breathing modes (RBM). Because Raman is a resonant process for SWNTs, different spectra corresponding to different resonant tubes are measured for each exciting laser line [6-9]. By using four wavelengths and measuring Stokes (S) and Anti-Stokes (AS) spectra, we excited the sample at eight different resonant energies. In the right part of Fig.3, we plot the sum of all spectra normalized to the total area of the RBM bands in both S and AS case. In order to determine nanotube diameter from Raman results, we used a phenomenological power law, validated by experiments and calculations [6]:

$$\nu_{cm^{-1}} = \frac{238}{d_{nm}^{0.93}} \quad (1)$$

Results are reported in Table 1. Remark that there is a minimum of intensity around 230 cm$^{-1}$ in the HiPCO spectra, which corresponds to a small amount of tubes of diameter around 1.05 Å. This "hole" may be due, in part, to a lack of adapted laser-line to cover this region. Indeed, *Kukovecz et al* measured Raman RBM on HiPCO samples using a laser line at 568.8 nm [7]. However, it can not be ruled out from their data that the distribution of diameters is bimodal, as also suggested by X-Ray data, with a minimum amount of 1.05 Å-diameter tubes.

Fig. 4 displays optical absorption experiments performed on the same samples. The spectra display several major absorption bands in the visible and near IR, due to optical transitions between pairs of van Hove singularities [9]. The allowed optical transitions have been calculated by Kataura *et al* as a function of diameter [9]. Note that the "Kataura plot" is also useful to analyze the resonant Raman results. From the position of the main absorption bands and using the Kataura plot, the tube diameter distribution can be estimated for each sample. Results are reported in Table 1. Note that the main broad bands are the envelopes of series of narrower bands, which are the signature of individual tubes of different chiralities, as expected from the "trigonal wrapping effect" [11]. These bands are better evidenced in the frequency range from 350 to 900 nm after substraction of an exponential background, as displayed in the inset. A detailed analysis of these absorption spectra will be reported elsewhere.

## DISCUSSION AND CONCLUSION

Table 1 compares the bundle and tube diameter for each sample, as estimated from SEM, HRTEM, XRD, Raman and optical absorption. There is a good qualitative agreement between all techniques. In EA and laser samples, SWNTs present a narrow diameter distribution centered around 1.35 nm. For HiPCO samples, the mean-diameter is shifted towards small values, around 1 nm, and the distribution is much broader but it is not clear, from our data, whether the distribution is monomodal or bimodal. The size of the bundles is also varying significantly from one sample to another: HiPCO SWNTs assemble into rather small bundles with respect to EA SWNTs. In the future, the possibility of working on different high-purity well-characterized samples and presenting different structural features opens interesting opportunities to precise the relation between structure and physical properties.

## ACKNOWLEDGMENTS

The authors acknowledge A. Loiseau and R. Almairac for fruitful discussions and for their help in HRTEM and XRD measurement.